\documentclass[
twocolumn,showpacs,preprintnumbers,amsmath,amssymb,eqsecnum]{revtex4}

\usepackage{graphicx}

\usepackage{bm}

\begin{document}

\title{Theory of elementary excitations in unstable Bose-Einstein
  condensates}

\author{U. Leonhardt$^{1}$}
\author{T. Kiss$^{1,2,3}$}
\author{P. \"Ohberg$^{1,4}$}

\affiliation{$^{1}$School of Physics and Astronomy, University of St
Andrews, North Haugh, St Andrews, KY16 9SS, Scotland }
\affiliation{$^{2}$Research Institute for Solid State Physics and Optics,\\ 
H-1525 Budapest, P.~O.~Box 49,  Hungary}
\affiliation{$^{3}$
Institute of Physics, University of P\'ecs,
Ifj\'us\'ag u.\ 6. H-7624 P\'ecs, Hungary}
\affiliation{$^{4}$Department of Physics, University of Strathclyde,\\
Glasgow G4 0NG, Scotland}      
             
\begin{abstract}
Like classical fluids, quantum gases may suffer from hydrodynamic instabilities.
Our paper develops a quantum version of the classical stability analysis in fluids, the Bogoliubov theory of elementary excitations in unstable Bose-Einstein condensates.
In unstable condensates the excitation modes have complex frequencies.
We derive the normalization conditions for unstable modes such that they 
can serve in a mode decomposition of the non-condensed component.
Furthermore, we develop approximative techniques to determine the spectrum and 
the mode functions.
Finally, we apply our theory to a sonic white hole
and find that the spectrum of unstable modes is
intrinsically discrete.
\end{abstract}


\maketitle

\section{Introduction}
Instabilities may haunt classical as well as quantum fluids. For
example, classical supersonic flows can trigger shock waves 
\cite{Courant,LL6}
or moving obstacles in Bose-Einstein condensates \cite{Dalfovo} can
shed vortex pairs \cite{Vortexpairs}. In fact, a dynamical instability
is at the heart of vortex nucleation in rotating condensates
\cite{Sinha}. In classical fluid mechanics \cite{LL6} the stability of a
solution of the equations of motion is treated using stability
analysis. Assuming a small perturbation of the solution, the equations
are linearized in the perturbation and the eigenvalues of the
linearized problem decide the fate of the solution.  Complex
eigenfrequencies with positive imaginary parts indicate
instabilities. In the theory of quantum fluids such as Bose-Einstein
condensates \cite{Dalfovo} the equations of motion are linearized
around the mean field to find the elementary excitations. The ground
state of the condensate is, almost by definition, stable, yet
macroscopic flows of condensed atoms may develop instabilities. Here
it is important to understand how to test for dynamical instabilities
and how unstable fluctuations evolve.

In this paper we elaborate a theory of elementary excitations in
unstable Bose-Einstein condensates. Our work is primarily inspired by
recent proposals \cite{Garay1,Garay2,Garay3,Laval,LKOHawking} 
for generating analogs of
black holes using transsonic condensates, but our theoretical concepts
and tools may certainly find applications in other situations as well
\cite{BoseNova}. Surprisingly little systematic work has been
published on elementary excitations in unstable condensates, to the
best of our knowledge, despite the fundamental nature of the subject.
Inspired by the treatment of instabilities in quantum fields \cite{QF}, 
Garay {\it et al.} studied a quantum theory of instabilities in Bose-Einstein
condensates in an appendix \cite{Garay2} and in a brief book
contribution \cite{Garay3}. 
Yurovsky \cite{Yurovsky} developed an alternative theory
of instabilities in quantum fluids.
Section II of our paper elaborates on
these ideas, starting from the basic concepts of elementary
excitations in dilute quantum gases \cite{Fetter}. We put forward an
economic notation that allows us to derive the theory with as little
technical effort as possible. Section III addresses two important
approximative methods to describe unstable excitations analytically.
We present a brief summary of the frequently applied acoustic
approximation and develop a version of the WKB approximation that can
be extended to complex frequencies and complex variables. In Section
IV we apply all the developed concepts and techniques to the
analysis of sonic horizons, demonstrating so their problem-solving potential.

\section{Elementary excitations}

\subsection{Fluctuation field}

Consider a stationary Bose-Einstein condensate of atoms with
short-range repulsive interactions. Following Fetter \cite{Fetter} we
describe the dynamics of the bosonic atom field $\hat \psi
(t,{ \bf x})$ using the grand-canonical Hamiltonian
\begin{equation}
  \label{eq:grand} 
  \hat H -\mu \hat N = \int \hat \psi^\dagger \left( -
  \frac{\hbar^2 \nabla^2}{2 m} + U - \mu + \textstyle{\frac{1}{2}}g\, 
  \hat\psi^\dagger \hat \psi \right) \hat \psi \, d^3 x \,.
\end{equation}
Here $\hat N$ abbreviates the total number of particles, $\int \hat
\psi^\dagger \hat \psi \, d^3 x$, a conserved quantity, and $\mu$
denotes a constant, the chemical potential. (Because of
particle-number conservation, both the Hamiltonian $\hat H$ and the
grand-canonical Hamiltonian (\ref{eq:grand}) are equivalent.) We
assume that most of the atoms constitute a Bose-Einstein condensate
with macroscopic wave function
\begin{equation} 
  \psi_0 = \sqrt{\rho_0} \, e^{i S_0} \, ,
\end{equation}
such that the deviations of $\hat \psi$ from the mean field $\psi_0$
form a quantum field $\hat \phi$ of small fluctuations,
\begin{equation}
  \hat \psi = \psi_0 + e^{i S_0} \hat \phi \, .
\end{equation}
We expand the grand-canonical Hamiltonian (\ref{eq:grand}) up to
quadratic order in $\hat \phi$, and get
\begin{eqnarray}
  \hat H - \mu \hat N &=& \hat H_0 + \hat H_1 + \hat H_2 \, ,\nonumber \\
  \hat H_0 &=& \int \psi_0^* 
  \left( 
    - \frac{\hbar^2 \nabla^2}{2 m} + U - \mu + \textstyle{\frac{1}{2}} g \rho_0 
  \right) 
  \psi_0 \, d^3 x \, , \nonumber \\
  \hat H_1 &=& \int \hat \phi^\dagger 
  \left( 
    - \frac{\hbar^2 \nabla^2}{2 m} + U - \mu +g \rho_0 
  \right) \psi_0 \, d^3 x \nonumber\\
  &&+ \textrm{H.\ c.} \, , \nonumber \\
  \hat H_2 &=& \int \left[ \hat \phi^\dagger \left( {\cal T} + U - \mu \right)
  \hat \phi \right. \nonumber \\
  && \left. \quad + \textstyle{\frac{1}{2}} g \rho_0 
  \left( 4 \hat \phi^\dagger \hat \phi
  + \hat \phi^{\dagger 2} + \hat \phi ^2 \right) \right] \, d^3 x 
\end{eqnarray}
with the kinetic term
\begin{equation}
  {\cal T} = \frac{m}{2} \left( \frac{\hbar \nabla}{i m} +{\bf u} \right)^2
\end{equation}
and the condensate flow
\begin{equation}
  {\bf u} = \frac{\hbar}{m} \nabla S_0 \, .
\end{equation}
Here $\hat H_0$ describes the energy of the condensate. The Hamiltonian 
$\hat H_1$ would displace the mean value of the fluctuations, 
when acting on $\hat \phi$, unless we impose
the stationary Gross-Pitaevskii equation
\begin{equation}
  \left( - \frac{\hbar ^2 \nabla^2}{2 m} + U - \mu - g |\psi_0|^2 
  \right) \psi_0 = 0 \, ,
\end{equation}
which minimizes also the Hamiltonian $\hat H_0$ such that
\begin{equation}
  \label{eq:h0}
  \hat H_0 = - \frac{1}{2} \int mc^2\rho_0\, d^3 x \, .
\end{equation}
Here, and throughout this paper, $c$ denotes the local speed of sound, 
defined by
\begin{equation}
  \label{eq:c}
  m c^2 = g \rho_0 \, .
\end{equation}
The quadratic Hamiltonian $\hat H_2$ generates the equations of motion
of the fluctuation field $\hat \phi$ (Bogoliubov-deGennes
equations).  We found it advantageous to deviate from the traditional
notation of condensate fluctuations \cite{Dalfovo} and to combine
$\hat \phi$ and $ \hat \phi^ \dagger$ in one spinor field
\begin{equation}
  \hat \varphi = 
  \left( 
    \begin{array}{c}
      \hat \phi\\
      \hat \phi^\dagger
    \end{array}
  \right) \, .
\end{equation}
Our spinor representation serves as a convenient shorthand notation, 
which does not refer to the spin of the atoms of course.
In terms of this Bogoliubov spinor $\hat \varphi$ the fluctuation
field evolves as
\begin{eqnarray}
  \label{eq:BdG}
  i \hbar \partial_t \hat \varphi &=& B \hat \varphi \, ,\nonumber \\
  B&=& \left( T +U-\mu+2 m c^2 \right) \sigma_z + i m c^2
  \sigma_y \, ,\nonumber \\
  T &=& \frac{m}{2} \left( \frac{\hbar \nabla}{i m} + {\bf u}
  \sigma_z \right)^2 \, ,
\end{eqnarray}
where $T$ describes the kinetic energy of the fluctuations. 
Throughout this paper we use the
Pauli matrices in their standard representation
\begin{equation}
  \sigma_x = \left(
    \begin{array}{cc}
      0& \,\,1\\ 
      1& \,\,0
    \end{array}
  \right) \, , \,
\sigma_y = \left(
    \begin{array}{cc}
      0& -i\\
      i& 0
    \end{array}
  \right) \, , \,
\sigma_z = \left(
    \begin{array}{cc}
      1& 0\\
      0& -1
    \end{array}
  \right) \, .
\end{equation}
The Bogoliubov-deGennes equation (\ref{eq:BdG}) is non-Hermitian,
because of the anti-Hermitian spinor-mixing term $i m c^2 \sigma_y$.
Therefore, the spectrum of elementary excitations is not necessarily
real. Dynamical instabilities may emerge. Finally, as a consequence of
Eqs. (\ref{eq:BdG}), we find
\begin{equation}
  \label{eq:h2}
  \hat H_2 = \frac{i \hbar}{2} \int \left( \hat \phi^\dagger
  (\partial_t \hat \phi)-(\partial_t \hat \phi^\dagger) \hat \phi
  \right) \, d^3 x \, ,
\end{equation}
an expression that we need in the mode expansion of the Hamiltonian
$\hat H_2$.

\subsection{Mode expansion}

As in standard field theories, we expand the Bogoliubov spinor $\hat
\varphi$ into modes. First we note that $\hat \varphi$ is invariant
under the conjugation
\begin{equation}
  \overline{\hat \varphi} \equiv \sigma_x \hat \varphi^\dagger \, .
\end{equation}
Consequently, $\hat \varphi$ must have the mode structure
\begin{equation}
  \label{eq:modes}
  \hat \varphi = \sum_\nu \left( w_\nu \hat a_\nu + \overline{
  w}_\nu \hat a_\nu^\dagger \right) \, .
\end{equation}
The spinor $w_\nu$ comprises Bogoliubov's familiar $u_\nu$ and
$v_\nu$ modes \cite{Dalfovo},
\begin{equation}
  \label{eq:w}
  w_\nu = \left(
    \begin{array}{c}
      u_\nu\\
      v_\nu
    \end{array}
  \right) \, ,
\end{equation}
and $\overline w_\nu$ denotes the conjugated Bogoliubov spinor,
\begin{equation}
  \overline w_\nu = \sigma_x w_\nu^* = 
  \left( 
    \begin{array}{c}
      v_\nu^*\\
      u_\nu^*
    \end{array}
  \right) \, ,
\end{equation}
such that \cite{Dalfovo}
\begin{equation}
  \hat \phi = \sum_\nu \left( u_\nu \hat a_\nu + v_\nu^* \hat
  a_\nu^\dagger \right) \, .
\end{equation}
The mode functions $w_\nu$ are subject to the Bogoliubov-deGennes
equation
\begin{equation}
  \label{eq:BdGw}
   i \hbar \partial_t w_\nu = B w_\nu
\end{equation}
that implies
\begin{equation}
  i \hbar \partial_t \overline w_\nu = B \overline w_{\nu} \, .
\end{equation}
In a field theory, modes are orthonormal with respect to an
invariant scalar product $(w_1,w_2)$ with
\begin{equation}
  \label{eq:inv}
  \partial_t (w_1, w_2) =0 \, ,
\end{equation}
in order
to guarantee that the $\hat a_\nu$ and $\hat a_\nu^\dagger$ are
annihilation and creation operators. Such a scalar product is
\begin{equation}
  \label{eq:scalar}
  (w_1,w_2) = \int w_1^\dagger \sigma_z w_2 \, d^3 x \,,
\end{equation}
where $w^\dagger$ abbreviates $w^{* T}$.  This scalar product is
time-invariant, because
\begin{equation}
  \label{eq:property}
  B^\dagger \sigma_z = \sigma_z B \, .
\end{equation}
The $\hat a_\nu$ and $\hat a_\nu^\dagger$ are Bose annihilation and
creation operators if we require that
\begin{eqnarray}
  (w_\nu,w_{\nu'}) &=& \int \left( u_\nu^* u_{\nu'} -
  v^*_{\nu'} v_\nu \right) \, d^3 x = \delta_{\nu \nu'} \,
  ,\nonumber \\
(\overline w_\nu, w_{\nu'})&=& \int \left( v_\nu u_{\nu'} - u_\nu v_{\nu'}
  \right) \, d^3 x  = 0 \, .
\end{eqnarray}
So far, the mode expansion (\ref{eq:modes}) is fairly general.
Now consider single-frequency modes,
\begin{equation}
  \label{eq:single}
  i \partial_t w_\nu = \Omega_\nu w_\nu \, .
\end{equation}
We obtain from the invariance (\ref{eq:inv}) of the scalar product
(\ref{eq:scalar})
\begin{equation}
  0 = i \partial_t (w_1,w_2)=(\Omega^*_2-\Omega_1)(w_1,w_2) \, .
\end{equation}
Consequently, single-frequency modes are only normalizable when their
frequencies $\Omega_\nu$ are real $\omega_\nu$ \cite{Fetter}. In this
case we obtain the well-known mode expansion of the Hamiltonian $\hat
H_2$ \cite{Fetter} in terms of independent harmonic oscillators,
\begin{equation}
  \hat H_2 = \sum_\nu \hbar \omega_\nu 
  \left( 
    \hat a_\nu^\dagger \hat a_\nu - 
    \int |v_\nu|^2 \, d^3 x 
  \right) \, .
\end{equation}
Single-frequency Bogoliubov spinors with complex $\Omega$, indicating
instabilities, have zero norm. However, this fact does of course not 
prohibit the existence of instabilities. It only means that we must not
directly employ such spinors as modes.

\subsection{Unstable condensates}

Suppose that the frequency of a Bogoliubov spinor $w$ is complex,
\begin{equation}
  \label{eq:omega}
  \Omega = \omega + i \gamma \, .
\end{equation}
Because the Bogoliubov-deGennes equation (\ref{eq:BdGw}) is
non-Hermitian, each eigenfrequency corresponds to a left and a right
eigenfunction, here denoted by $w_+$ and $w_l$, respectively,
\begin{equation}
  B w_+ = \hbar \Omega\, w_+ \, , \quad
  w_l^\dagger B = \hbar \Omega\, w_l^\dagger \, .
\end{equation}
As a consequence of the property (\ref{eq:property}) we find
\begin{equation}
B (\sigma_z w_l) = 
\sigma_z B^\dagger w_l = 
\hbar \Omega^* (\sigma_z w_l) \, .
\end{equation}
\vspace{1mm}

\noindent
Therefore, $\sigma_z w_l$ is the Bogoliubov spinor with the complex
conjugated eigenfrequency of $w_+$. Complex frequencies of elementary
excitations occur in conjugated pairs, reflecting the Hermiticity of the 
grand-canonical Hamiltonian (\ref{eq:grand}). 
The spinor conjugate $\overline w_+$ corresponds trivially to the 
frequency $-\Omega^*$, whereas the
frequency $-\Omega$ is associated with the spinor
\begin{equation}
  w_- \equiv \overline{\sigma_z w_l} \, .
\end{equation}
Consider the scalar product of the modes with $\pm \Omega_n$ frequencies 
that are labeled by the subscripts $\pm n$,
\begin{equation}
  0=i\partial_t (\overline w_{-n}, w_{+n'})=(-\Omega_n+\Omega_{n'})(\overline
  w_{-n}, w_{+n'}) \, .
\end{equation}
Consequently, we can require
\begin{equation}
  \label{eq:ortho}
  (\overline w_{-n}, w_{+n'}) = \delta_{n n'}
\end{equation}
by choosing the appropriate overlap between the left and right
eigenstates of $B$. In Section IV we use this orthogonality condition
to find the unstable elementary excitation of a sonic horizon.

Single-frequency Bogoliubov spinors with complex $\Omega$ must not
represent modes {\it per se}, yet nothing prevents us from combining
two or more of such spinors to form non-stationary modes. A simple
choice is
\begin{equation}
  {\cal W}_{\pm n} \equiv 
  \frac{1}{\sqrt{2}} (w_{\pm n}\pm \overline w_{\mp n}) =
  \left(
    \begin{array}{c}
      {\cal U}_{\pm n}\\
      {\cal V}_{\pm n}
    \end{array}
    \right)
\end{equation}
satisfying the relations
\begin{eqnarray}
  \label{eq:onr}
  ({\cal W}_{\pm n},{\cal W}_{\pm n'}) &=& \delta_{n n'}\, , \,
  ({\cal W}_{\pm n},{\cal W}_{\mp n'}) = 0\nonumber \\
  (\overline {\cal W}_{\pm n},{\cal W}_{\pm n'}) &=& 0\, , \quad\, \,
  (\overline {\cal W}_{\pm n},{\cal W}_{\mp n'}) = 0 \,.
\end{eqnarray}
Therefore, the ${\cal W}_{\pm n}$ are perfectly suitable as Bogoliubov
modes. We expand the fluctuation field $\hat \phi$ in terms of the
${\cal U}_\pm, {\cal V}_\pm$ modes, 
\begin{eqnarray}
  \label{eq:exp}
  \hat \phi &=& \sum_\nu \hat \phi_n \, ,\nonumber \\
  \hat \phi_n &=& 
  {\cal U}_{+n} \hat a_{+n} + 
  {\cal V}_{+n}^* \hat a_{+n}^\dagger +
  {\cal U}_{-n} \hat a_{-n} + 
  {\cal V}_{-n}^* \hat a_{-n}^\dagger \, .
\end{eqnarray}
We obtain from Eqs. (\ref{eq:single}) and (\ref{eq:omega})
\begin{eqnarray}
  i \partial_t {\cal U}_{\pm n} &=&  
  \pm \omega_n {\cal U}_{\pm n} - i \gamma_n {\cal V}_{\mp n}^* \, , \,
  \nonumber\\
  i \partial_t {\cal V}_{\pm n} &=& 
  \pm \omega_n {\cal V}_{\pm n}- i \gamma_n {\cal U}^*_{\mp n}\, , \,
\end{eqnarray}
and, consequently,
\begin{widetext}
\begin{equation}
  i \partial_t \hat \phi_n =
  \left(
    \omega_n \hat a_{+n} - i \gamma_n \hat a_{-n}^\dagger 
  \right) 
  {\cal U}_{+n}
   - \left(
    \omega_n \hat a_{+n}^\dagger + i \gamma_n \hat a_{-n} 
  \right) 
  {\cal V}_{+n}^* 
  - \left(
    \omega_n \hat a_{-n} + i \gamma_n \hat a_{+n}^\dagger 
  \right) 
  {\cal U}_{-n}
  + \left(
    \omega_n \hat a_{-n}^\dagger - i \gamma_n \hat a_{+n} 
  \right) 
  {\cal V}_{-n}^* \,.
\end{equation}
We insert this result and the expansion (\ref{eq:exp}) into formula
(\ref{eq:h2}), use the orthonormality relations (\ref{eq:onr}), and
get
\begin{eqnarray}
  \hat H_2 &=& \hbar \sum_n  \omega_n \Big[ \hat a_{+n}^\dagger \hat
  a_{+n} - \hat a_{-n}^\dagger \hat a_{-n}
   - \int \left( |{\cal
  V}_{+n}|^2 -|{\cal V}_{-n}|^2\right) \, d^3 x \Big]
  \nonumber\\
  && + \hbar \sum_n i \gamma_n \Big[ \hat a_{+n} \hat a_{-n} - \hat
  a_{+n}^\dagger \hat a_{-n}^\dagger
  + \int \left( {\cal U}_{+n} {\cal
  V}_{-n} - {\cal U}_{+n}^* {\cal V}_{-n}^* \right) \, d^3 x
  \Big] \, .
\end{eqnarray}
\end{widetext}
Due to the instability of the condensate, pairs of elementary
excitations are spontaneously generated at the rates $\gamma_n$, 
and so the non-condensed part grows at
the expense of the condensate. Of course, the Hamiltonian $\hat H_2$
describes the correct dynamics only for short times, as long as the
growing excitations are still small compared with the condensate.
Furthermore, the backaction of the non-condensed part onto the
condensate ought to be taken into account, affecting the growth rates
$\gamma_n$ and the frequencies $\omega_n$. 
The instability causes the condensate to dissolve.
Nevertheless, the atoms may settle afterwards to constitute a new
condensate with a stable macroscopic wave function, as happens in
vortex nucleation \cite{Sinha}.

\section{Approximative methods}

\subsection{Acoustic approximation}

Frequently, approximative methods provide the tools to find analytic
results that capture the essential physics of elementary excitations.
The best known example is the excitation spectrum of a condensate in a
harmonic trap \cite{StringariOhberg}. Here the excitations of the
condensate have been calculated in hydrodynamic or, as we would prefer 
to call it, acoustic approximation. (Elementary excitations are sound waves
within the validity of the approximation.) Furthermore, sound waves in
moving Bose-Einstein condensates propagate in the same way as
mass-less waves in a sufficiently large class of curved space-time
structures \cite{Garay1,Garay2,Garay3,Unruh}. In Section IV we use
this connection to analyze the instabilities of a sonic horizon.

Let us briefly summarize the main aspects of the acoustic
approximation. Given a solution (\ref{eq:w}) of the Bogoliubov-deGennes 
equation (\ref{eq:BdGw}), the function
\begin{equation}
  \label{eq:psi1}
  \psi=\psi_0 + e^{i S_0}(u+v^*)
\end{equation}
solves the time-dependent Gross-Pitaevskii equation,
\begin{equation}
  \label{eq:GPE}
  i\hbar\partial_t\psi=
  \left( - \frac{\hbar ^2 \nabla^2}{2 m} + U - g |\psi_0|^2
  \right) \psi \, ,
\end{equation}
as long as $u$ and $v$ are small. 
We represent $\psi$ in terms of the particle density $\rho$
and the phase $S$,
\begin{equation}
  \label{eq:psi2}
  \psi = \sqrt{\rho}\, e^{i S} \, , \quad \rho = \rho_0+\rho_s
  \, , \quad S=S_0+s \, .
\end{equation}
Neglecting the quantum potential $\hbar^2 (\nabla^2 \sqrt{\rho})/(2
m \sqrt{\rho})$  in the Gross-Pitaevskii equation (\ref{eq:GPE}),
we recover the equation of continuity and the Bernoulli equation,
\begin{eqnarray}
  \label{eq:hydro}
  \partial_t \rho + \nabla \cdot \left( \rho \frac{\hbar}{m} \nabla S
  \right) &=& 0 \, ,\\
  \hbar \partial_t S + \frac{\hbar^2}{2 m} (\nabla S)^2 + g \rho +
  U &=& 0 \, .
\end{eqnarray}
Assuming that $\rho_0$ and $S_0$ satisfy Eq. (\ref{eq:hydro}) and
linearizing in $\rho_s$ and $s$ gives
\begin{eqnarray}
  \partial_t \rho_s &=& -\nabla \cdot \left( \rho_s {\bf u} + \rho_0
  \frac{\hbar}{m} \nabla s \right) \, ,\\ 
  \label{eq:bern}
  \rho_s &=& - \frac{\hbar\rho_0}{m c^2} 
  (\partial_t + {\bf u} \cdot \nabla) s \, ,
\end{eqnarray}
where $c$ denotes the speed of sound (\ref{eq:c}) and
${\bf u}$ describes the flow $(\hbar/m) \nabla S_0$. Substituting the
expression for $\rho_s$ produces the wave equation for sound
in irrotational fluids \cite{Unruh, Stone}
\begin{eqnarray}
  \label{eq:waves}
  \partial_\mu f^{\mu \nu} \partial_\nu s &=& 0 \, , \nonumber \\
  f^{\mu \nu} &=& \frac{\rho_0}{c^2}\left(
    \begin{array}{cc}
      1&\,\,{\bf u} \\
      {\bf u}&\,\,-c^2 \openone + {\bf u} \otimes{\bf u} 
    \end{array}
  \right)\, , 
  \nonumber\\
  \partial_\nu &=& (\partial_t, \nabla) \, ,
\end{eqnarray}
the central argument in the analogy between sound in moving media and
waves in general relativity \cite{Garay3,Unruh}.  Here we have used a
relativistic notation with $\mu$ and $\nu$ referring to space-time
coordinates (not to the chemical potential of course).

To see how $\rho_s$ and $s$ are related to the Bogoliubov spinor we
compare $\psi^2$ of Eqs. (\ref{eq:psi1}) and (\ref{eq:psi2}) to linear
order in $\rho_s, \, s, \, u$ and $v$, and get
\begin{equation}
   \label{eq:rep}
  u+v^* = \sqrt{\rho_0} 
  \left( 
    \frac{ \rho_s}{2 \rho_0} + i s
  \right) \, .
\end{equation}
Consequently, a single-frequency sound wave appears as the excitation
mode
\begin{eqnarray}
  \label{eq:relation}
  u &=& \sqrt{\rho_0} \left( \frac{\hbar}{2m c^2} 
    ( -i\Omega + {\bf u}\cdot\nabla) + 1 \right) 
  i\sigma e^{-i \Omega t} \, , \nonumber \\
  v &=& \sqrt{\rho_0} \left( \frac{\hbar}{2m c^2} 
    (-i\Omega + {\bf u}\cdot\nabla) - 1 \right) 
  i\sigma e^{-i \Omega t}
\end{eqnarray}
where
\begin{equation}
  s= \sigma e^{-i \Omega t} + \sigma^* e^{+i \Omega^* t} \, .
\end{equation}

\subsection{WKB approximation}

Another important approximative method to analyze elementary
excitations is the WKB approximation \cite{Csordas}, the equivalent of
semiclassical wave mechanics or geometrical optics (geometrical
acoustics). Frequently, the important features of waves are determined
by the turning points of rays on the complex plane or by branch points
of the momentum. Here we develop a modification of the WKB
approximation for elementary excitations \cite{Csordas} that can be
analytically continued to complex variables and complex frequencies.

We assume an effectively one-dimensional model with spatial coordinate
$z$ and flow $u$. We express the Bogoliubov spinor as
\begin{equation}
  \label{eq:ansatz}
  w= (w_0+ \hbar w_1 + \dots) \exp 
  \left[ 
    \frac{i}{\hbar}\left( \int p \, dz - E t \right)
  \right]  
\end{equation}
in terms of the semiclassical momentum $p$ and the energy $E$. 
We insert the ansatz
(\ref{eq:ansatz}) into the Bogoliubov-deGennes equation
(\ref{eq:BdGw}) and expand the result into powers in $\hbar$. In
zeroth order we obtain
\begin{eqnarray}
  \label{eq:zeroth}
  B_0 w_0 &=& E w_0 \,, \nonumber \\
  B_0 &=& \left( \frac{1}{2 m} ( p \openone+ m u \sigma_z)^2 +(U-\mu + 2
  m c^2) \right) \sigma_z \nonumber\\
  && \quad + i m c^2 \sigma_y \, .
\end{eqnarray}
The determinant of $B_0- E \openone $ vanishes when $p$ satisfies the
Hamilton-Jacobi equation
\begin{equation}
  \label{eq:HJE}
  \left( \frac{p^2}{2 m} + E_0 \right)^2 -(E-up)^2 =m^2 c^4
\end{equation}
with
\begin{equation}
  E_0 = U+\frac{m}{2} u^2 - \mu + 2 m c^2 \, .
\end{equation}
The components $u_0$ and $v_0$ of the envelope $w_0$ are linearly 
dependent, because $w_0$ is an eigenstate of $B_0$. We get
\begin{eqnarray}
  \label{eq:eta}
  v_0 &=&  \eta u_0 \, , \nonumber \\
  \eta &=& - \frac{1}{m c^2} \left( E_0+\frac{p^2}{2 m} + u p - E
  \right) \, .
\end{eqnarray}

Before we proceed,
let us see how the Hamilton-Jacobi equation is related to the
acoustic approximation. The hydrodynamical model (\ref{eq:hydro})
presumes small local variations in the density and in the flow such
that we can describe the condensate locally. Let us assume that the
chemical potential $\mu$ corresponds reasonably accurately to the local
energy of the condensate, $m c^2 +U + \frac{m}{2} u^2$, such that
\begin{equation}
  \label{eq:ueff}
  E_0 = m c^2 \, .
\end{equation}
Then, for $|p|\ll mc$, {\it i.e.} for sound with wavelengths
much larger than the healing length $\hbar/(\sqrt{2} m c)$
\cite{Dalfovo}, we get the dispersion relation of sound in moving media
\begin{equation}
  (E-up)^2 = c^2 p^2 \, .
\end{equation}
Additionally, in the relationship  (\ref{eq:eta}) between the envelope
components  we ignore the $p^2/(2m)$ term within the acoustic 
approximation. We get
\begin{equation}
\eta = -1 + \frac{E-up}{mc^2} = -1 \pm \frac{p}{mc} \,,
\end{equation}
which represents the linearized Bernoulli equation (\ref{eq:bern})
expressed in terms (\ref{eq:rep}) of the Bogoliubov-spinor components.

To first order in $\hbar$ we obtain from the Bogoliubov-deGennes
equation (\ref{eq:BdGw}) and from the ansatz (\ref{eq:ansatz})
\begin{eqnarray}
  0&=&B_1 w_0 + \sigma_z (B_0 - E \openone ) w_1  \, , \nonumber \\
  B_1 &=& \frac{1}{2 i m} \Big( ( 2 p \partial_z + p') \openone + m ( 2
  u \partial_z + u') \sigma_z \Big) \, .
\end{eqnarray}
For complex frequencies or complex $z$ values the matrix $\sigma_z
(B_0 - E \openone)$ is symmetric but not necessarily Hermitian. Since
$w_0$ is the eigenvector of $\sigma_z(B_0 - E \openone)$ with zero
eigenvalue, $B_1w_0$ must be orthogonal on $w_0$ with respect to the
scalar product $w_1^T w_2$. Note that this scalar product does not
involve complex conjugation. We find
\begin{eqnarray}
  \label{eq:cont}
  0 &=& w_0^T B_1 w_0 \nonumber \\
  &=& \frac{1}{2 i m} \partial_z \Big( (u_0^2+v_0^2)p + m (u_0^2 -
  v_0^2) u \Big) \, ,
\end{eqnarray}
which gives the continuity relation
\begin{equation}
  \label{eq:continuity}
  \partial_z ( u_0^2 - v_0^2) v = 0 
\end{equation}
with the velocity
\begin{equation}
  v= \frac{1+\eta^2}{1-\eta^2} \frac{p}{m} + u = \frac{ \partial
  H}{\partial p}
\end{equation}
in terms of the semiclassical Hamiltonian
\begin{equation}
  \label{eq:ham}
  E=H= u p \pm \sqrt{ \left( \frac{p^2}{2 m} + E_0 \right)^2 - m^2
  c^4} \, .
\end{equation}
The continuity relation (\ref{eq:continuity}) shows how the spinor
amplitudes are connected on the complex plane.  For real energies and
real coordinates $(|u_0|^2-|v_0|^2)v$ is exactly conserved, describing a
stationary quasiparticle flux \cite{Csordas}. In this case our result 
agrees with $(|u_0|^2-|v_0|^2)v$, apart from an arbitrary phase.
On the other hand, for complex energies or complex coordinates 
$(u_0^2 -v_0^2)v$ does not correspond to an exact conservation law 
in general, but still remains a constant within the validity 
of the WKB approximation.
The advantage of our result is that it can be analytically continued
such that
\begin{equation}
  u_0^2-v_0^2=\frac{A_0}{v} \,,
\end{equation}
with a constant $A_0$
as long as $z$ does not reach the vicinity of a turning point or
crosses a Stokes line in the complex plane \cite{Stokes}.

\subsection{Turning points}

At a turning point the velocity $v$ vanishes and, consequently,
$u_0^2-v_0^2$ diverges such that the WKB approximation is no longer
valid in its vicinity. Turning points are the origins of Stokes lines
where the WKB solutions are discontinuously connected \cite{Stokes}.

Assume that the condensate is nearly uniform around a turning point
such that the speed of sound is approximately constant and that the
locality condition (\ref{eq:ueff}) is satisfied. In this case the
turning point does not lie at the edge of the condensate, as it is the
case for oscillations in harmonic traps \cite{StringariOhberg}. Let us
find out whether and where such turning points exist. Our conditions
imply that the momentum $p$ depends on $z$ only through the flow $u(z)$. 
Let us turn matters around and regard $z$ as a function of $u$ and $u$ 
as a function of $p$. We get from the expression (\ref{eq:ham}) of the
quasiclassical Hamiltonian
\begin{equation}
  v= \frac{\partial H}{\partial p} = u + \frac{\partial}{\partial p} (
  E - up ) = - p \frac{\partial u }{\partial p} \, .
\end{equation}
The definition (\ref{eq:eta}) of $\eta$ and the
Hamilton-Jacobi equation (\ref{eq:HJE}) 
with the condition (\ref{eq:ueff}) implies the relation
\begin{equation}
  \label{eq:rel}
  \frac{p^2}{m^2 c^2} = - \frac{(1+\eta)^2}{\eta} \, .
\end{equation}
We substitute  this result for $p^2$ in Eq. (\ref{eq:eta}), solve for $up$, 
and get
\begin{equation}
  \label{eq:uuu}
  u = \frac{m c^2}{p} \left( \varepsilon - \frac{ \eta^2 -1}{2 \eta} \right)
\end{equation}
with
\begin{equation}
  \label{eq:eps}
  \varepsilon = \frac{E}{m c^2} \, .
\end{equation}
We differentiate Eq.\ (\ref{eq:uuu}) with respect to the momentum 
$p$, utilize relation (\ref{eq:rel}) and its momentum derivative,
and arrive at
\begin{equation}
  \label{eq:dup}
  \frac{\partial u}{\partial p} = \frac{1+(3-2 \varepsilon)
  \eta + (3+2 \varepsilon) \eta^2 + \eta ^3}{2 m (\eta-1)(\eta+1)^2} \, .
\end{equation}
The turning points correspond to the zeros $\eta_0$ of the numerator.
We note that $ \varepsilon$ is naturally a small number, 
because the energy of the elementary excitation ought to be much 
smaller than the condensate's energy (\ref{eq:h0}). 
We expand $\eta_0$ in powers of $\varepsilon^{1/3}$,
\begin{equation}
  \label{eq:series}
  \eta_0 = x_0 + x_1 \varepsilon^{1/3}+ x_2
  \varepsilon^{2/3} + x_3 \varepsilon^{3/3} + \dots \, ,
\end{equation}
and find the coefficients
\begin{equation}
  x_0=-1 \, , \,\, x_1 = \sqrt[3]{-4} \, , \,\, x_2 = \frac{2}{x_1} \, , \,\,
  x_3 = - \frac{2}{3} \, .
\end{equation}
Close to a turning point the flow profile depends quadratically 
on the momentum
\begin{equation}
  \label{eq:branches}
  u-u_0 =\left. \frac{\partial^2 u}{\partial p^2}\right|_0
  \frac{(p-p_0)^2}{2} \,.
\end{equation}
This relation shows that a turning point is a branch point for the 
semiclassical momentum $p$ and it also specifies the onset of 
Stokes lines \cite{Stokes},
defined as the lines where the differences 
between the phases $\int p\,dz$ of the two $p$ branches is purely 
imaginary. Here one of the waves is exponentially small 
compared with the other. Crossing a Stokes line connects waves
in a  discontinuous yet precisely defined way \cite{Stokes}.
We obtain the coefficients from Eqs.\ (\ref{eq:rel}) 
and (\ref{eq:dup}), using the series (\ref{eq:series}),
\begin{eqnarray}
  \pm \frac{p_0}{mc} &=& 
   x_1\varepsilon^{1/3} -\frac{\varepsilon}{6} + \dots \,, \nonumber\\
   \left. \pm \frac{\partial^2 u}{\partial p^2}\right|_0 &=&
   -\frac{3}{4}-\frac{5}{8x_1}\,\varepsilon^{2/3}-
   \frac{\varepsilon}{4} + \dots \,.
   \label{eq:ddur}
\end{eqnarray}
Finally, we determine the velocity 
$u_0$ at a turning point from Eq.\ (\ref{eq:uuu}). 
We find
\begin{equation}
  \pm \frac{u_0}{c} = -1 + \frac{3}{2} \sqrt[3]{-1} \left(
  \frac{\varepsilon}{2}\right)^{2/3} + {\rm O}(\varepsilon^{4/3})  \,.
\end{equation}
Here $\sqrt[3]{-1}$ refers to the three cubic roots of $-1$,
generating three turning points in the complex plane when $u$
approaches $c$. In a mostly uniform condensate the turning points of
elementary excitation are close to transsonic regions where the
condensate flow transcends the speed of sound. Such a region forms a
sonic horizon.

\section{Sonic horizons}

\subsection{Model}

Consider a stationary transsonic medium {\it i.e.}\ a medium with a
spatially non-uniform flow that varies from subsonic to supersonic
speed. One would expect that beyond the interface where the flow exceeds 
$c$ sound waves are swept away such that no sound from the supersonic 
zone can return to the subsonic region. 
This transsonic interface serves as the
sonic equivalent of a black-hole horizon.
\cite{Garay1,Garay2,Garay3,Unruh}. 
On the other hand, the interface where the flow settles from supersonic 
to subsonic speed forms the horizon of a sonic white hole 
\cite{Garay1,Garay2,Garay3}, 
an object that no sound wave can enter from outside. 
Close to the horizons, 
sound waves propagating against the current freeze, and their
wavelengths are dramatically reduced.

Transsonic Bose-Einstein condensates offer great prospects
\cite{Garay1,Garay2,Garay3,Laval,LKOHawking} 
for demonstrating the quantum effects of
event horizons \cite{Hawking,Brout}. Such effects and the stability of
the condensate depend on the behaviour of the condensate close to the
horizons. Let us thus focus on the physics in the vicinity of a 
black- or white-hole horizon. 
In this case we can use the simple one-dimensional model
\begin{equation}
  \label{eq:linear}
  u = -c + \alpha z \, .
\end{equation}
Here $z$ denotes the spatial coordinate orthogonal to the horizon at
$z=0$, $\alpha$ characterizes the surface gravity or, in our
acoustic analog, the gradient of the transsonic flow, and
$\rho_0$ and $c$ are assumed to be constant. 
Strictly speaking, we should complement the flow component 
(\ref{eq:linear}) in the $z$ direction by appropriate components 
in the $x$ and $y$ directions,
in order to obey the continuity of the flow.
But as long as we focus on effects on length scales smaller than
$|c/\alpha|$ we can ignore the other dimensions of the fluid.
Depending on the sign of $\alpha$, two cases emerge.
When the velocity gradient is positive we are considering
the horizon of a sonic black hole.
When $\alpha$ is negative the horizon refers to a white hole
\cite{Garay1,Garay2,Garay3}.
In a typical alkali Bose-Einstein condensate 
without exploitation of Feshbach resonances 
the speed of sound $c$ is in the order of $1\, {\rm mm}/{\rm s}$.
The transsonic velocity gradient should be small compared 
with the healing length \cite{Dalfovo}
\begin{equation}
 |\alpha|\,\xi \ll c \,,\quad \xi = \frac{\hbar}{mc\sqrt{2}} \,,
\end{equation}
which guarantees that the Hawking energy $k_{_B} T_{_H}$
is much smaller than the energy of the condensate,
\begin{equation}
 k_{_B} T_{_H}=\frac{\hbar|\alpha|}{2\pi} \ll mc^2  \,.
\end{equation}
Being in the vicinity of the horizon and having the linear velocity
profile (\ref{eq:linear}) presumes that
\begin{equation}
  |z| \ll \left|\frac{c}{\alpha}\right| \, .
\end{equation}
The energies of elementary excitations should be sufficiently smaller 
than $mc^2$,
which implies that their dimensionless energy parameter $\varepsilon$ 
defined in Eq.\ (\ref{eq:eps}) is small.
The excitations are sound waves for low wavenumbers and for
$z$ far away from the turning points,
\begin{equation}
  \label{eq:turn}
  |z| \gg |z_0| \, , \quad z_0 = \frac{3c}{2\alpha} \sqrt[3]{-1}
   \left( \frac{\varepsilon}{2} \right)^{2/3} \, .
\end{equation}
Given the linear velocity profile (\ref{eq:linear}), we solve the wave
equation (\ref{eq:waves}) exactly, and get
\begin{equation}
  s = {s_0}\left(i z^{\pm i \Omega/\alpha}e^{-i
  \Omega t} - i z^{\mp i \Omega^*/\alpha} e^{i \Omega^*t} \right) \, .
\end{equation}
We use the relationship (\ref{eq:relation}) to find the Bogoliubov
spinor within the acoustic approximation,
\begin{equation}
 \label{eq:wa}
  w = A \left(
    \begin{array}{c}
      \displaystyle\frac{\Omega}{2 \alpha z} + \frac{mc}{\hbar}\\
      \displaystyle\frac{\Omega}{2 \alpha z} - \frac{mc}{\hbar}  \end{array}
\right)
z^{\pm i \Omega/\alpha}e^{-i\Omega t} \, ,
\end{equation}
where 
\begin{equation}
A=-s_0\frac{\hbar\sqrt{\rho_0}}{mc}\,.
\end{equation}
The approximation (\ref{eq:wa})
is restricted to complex $z$ variables outside the
trans-acoustic region indicated in Fig.\ 1.
\begin{figure}
\includegraphics[width=7cm]{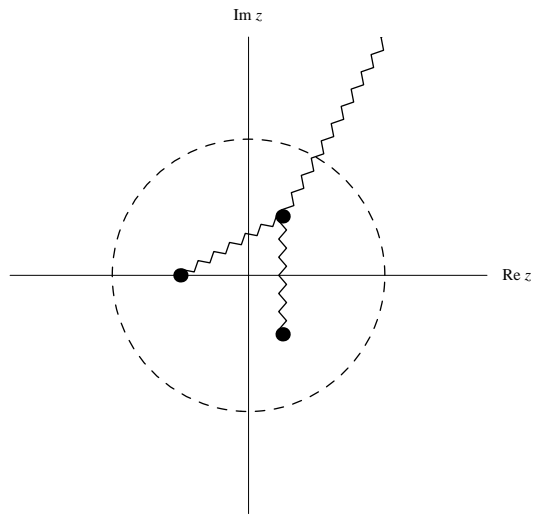}
\caption{\label{fig:complex} 
Analytic structure of a sonic horizon in a 
Bose-Einstein condensate. 
The figure shows the three turning points in the
complex $z$ plane around the horizon at the origin.
The points are connected by two branch cuts 
where the wavenumber $k$
of acoustic elementary excitations is elevated to other solutions
of the dispersion relation. 
The circle around the turning points roughly indicates the place
where $k$ approaches the asymptotics $\Omega/(\alpha z)$ 
that is characteristic of waves at a horizon \cite{Brout}.
The third branch cut, connecting one of the turning points
to infinity, is the cut between the two trans-acoustic branches 
with the asymptotics (4.11).}
\end{figure}
\begin{figure*}
\includegraphics{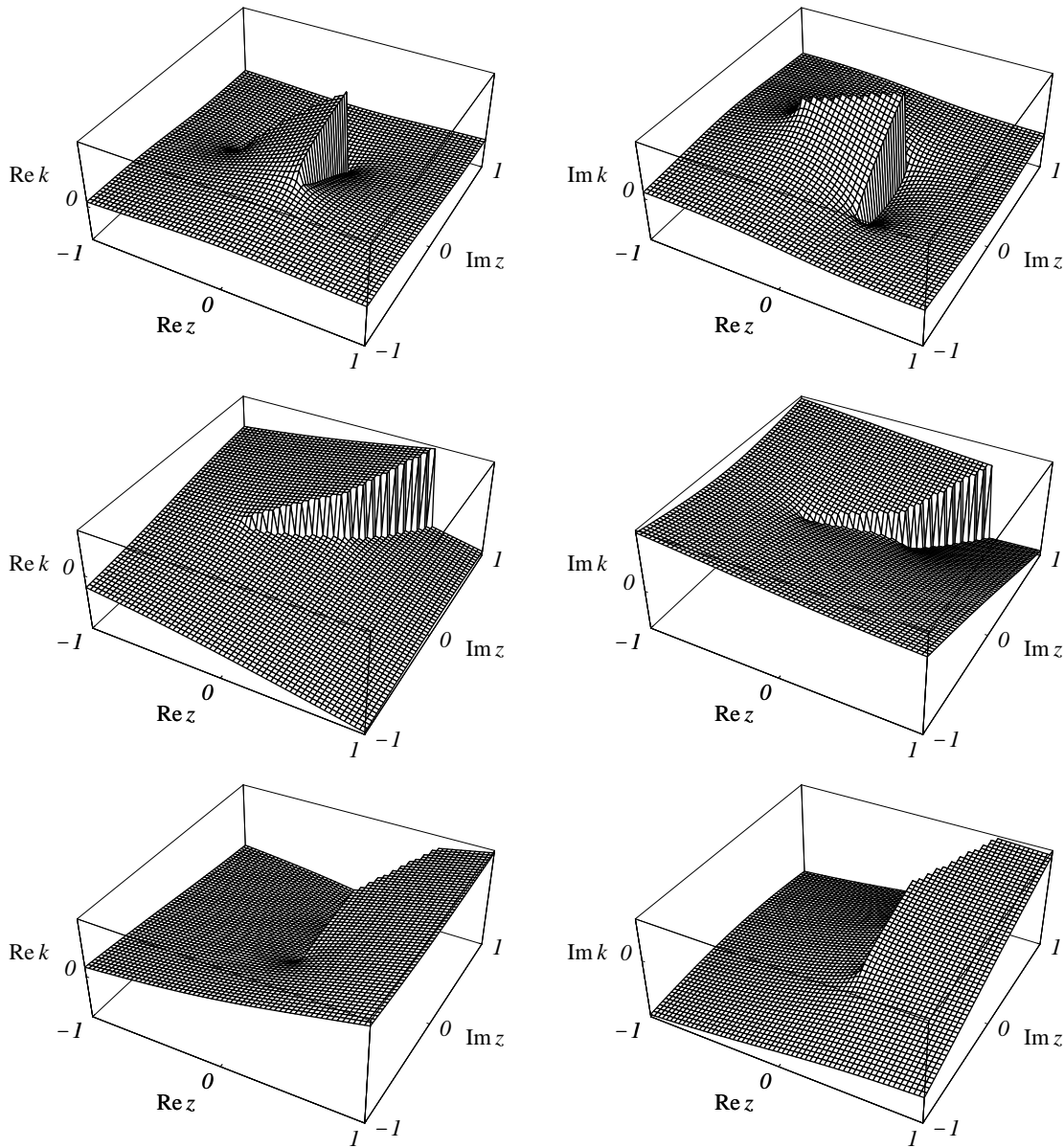}
\caption{\label{fig:branches}
Wavenumbers $k$ of elementary excitations 
around a sonic white-hole horizon, 
analytically continued on the complex plane.
The figure shows three roots of the dispersion relation
$[\hbar^2k^2/(2m) + mc^2]^2 -
\hbar^2[\Omega-(-c+\alpha z)k]^2 =m^2c^4$ for
$\Omega=0.1i \,(mc^2/\hbar)$ 
and $\alpha=-0.5 \,(mc^2/\hbar)$, 
illustrating the branch cuts of $k$.
The top row displays the wavenumber of a sound wave 
that propagates against the current. 
The picture indicates the characteristic $\Omega/(\alpha z)$ 
asymptotics away from the branch points. 
The two lower rows display two 
trans-acoustic branches of $k$. 
The fourth root of the dispersion relation is not shown,
because it corresponds to the trivial case of
sound waves that propagate with the flow.}
\end{figure*}
Close to the trans-acoustic zone the wavelength of elementary
excitations is dramatically reduced.
In this regime we can use the WKB approximation.
In the region where both the acoustic and the WKB approximation
are applicable we represent the Bogoliubov spinor as
\begin{equation}
  w = \left( 
    \begin{array}{c}
      u_0\\
      v_0
    \end{array}
  \right) \exp \left( i \int k \, dz - i \Omega t \right)
\end{equation}
with the wavenumber
\begin{equation}
  k \sim \frac{\Omega}{\alpha z} \, .
\end{equation}
If the acoustic approximation were universally valid the wavenumber
$k$ would approach a singularity at the horizon where, consequently,
the wavelength of sound would shrink to zero.  Apparently, Nature
tends to prevent such extreme behaviour \cite{TransPlanck}. In our
case, the singularity is split into three branch points, the turning
points. The branch cuts between the points connect the acoustic
branch to two other trans-acoustic branches with wavenumbers so
high that the branches do not represent sound waves subject to the
wave equation (\ref{eq:waves}).  We solve the Hamilton-Jacobi equation
(\ref{eq:HJE}) in the $|\Omega| \ll |uk|$ limit and get the
asymptotics for the trans-acoustic branches \cite{Corley}
\begin{equation}
  k \sim \pm 2 \frac{mc}{\hbar} \sqrt{u^2/c^2 - 1} + \frac{ \Omega u}{ c^2-u^2} \, .
\end{equation}
Figure 2 illustrates the momenta $k$ of the three branches in the
complex $z$ plane. Finally, we note that the fourth solution of the
Hamilton-Jacobi equation (\ref{eq:HJE}) corresponds to sound waves
that are swept away by the current, a less interesting case.

\subsection{Stokes phenomenon}

It is tempting to assume that we could employ the acoustic asymptotics 
$\Omega/(\alpha z)$ of the wavenumber $k$ on the entire complex plane,
as long as $z$ is sufficiently far away from the turning points at the horizon. 
In this case, however, the phase $\int k\,dz$ becomes logarithmic 
and hence multivalued, whereas the true Bogoliubov mode function
is singlevalued.
The Stokes phenomenon \cite{Stokes} resolves this conflict
by connecting the acoustic modes to trans-acoustic ones on the upper
or the lower half plane, 
giving rise to connection formulas  \cite{Corley}
that describe mode conversion \cite{CJ}.
Each turning point $z_0$ connects two branches of WKB wavenumbers,
as we see from Eq.\ (\ref{eq:branches}), and each $z_0$ is origin of
three Stokes lines.
At a Stokes line the phase difference between
the two connected branches is purely imaginary. 
Consequently, one of the WKB waves exponentially exceeds the other
such that the smaller cannot be resolved within the WKB approximation,
if the larger wave is present. 
In general, the Bogoliubov spinor is a superposition of the four 
fundamental solutions that correspond to the four branches of the WKB
wavenumbers.
\begin{equation}
  w = c_{_A}w_{_A} + c_{_B}w_{_B} + c_{_C}w_{_C} + c_{_D}w_{_D}
  \,.
\end{equation}
When crossing a Stokes line, the exponentially suppressed solution 
may gain an additional component that is proportional to the coefficient
of the exponentially enhanced solution. 
If we wish to construct a Bogoliubov spinor where only the exponentially
smaller component exits in the vicinity of a Stokes line 
we must put the coefficient of the larger one to zero. 
Figure 3 shows the Stokes lines of Boboliubov modes at
a sonic white hole with purely imaginary 
frequency, a case of importance in Sec.\ IV~C.
\begin{figure}
\includegraphics[width=7cm]{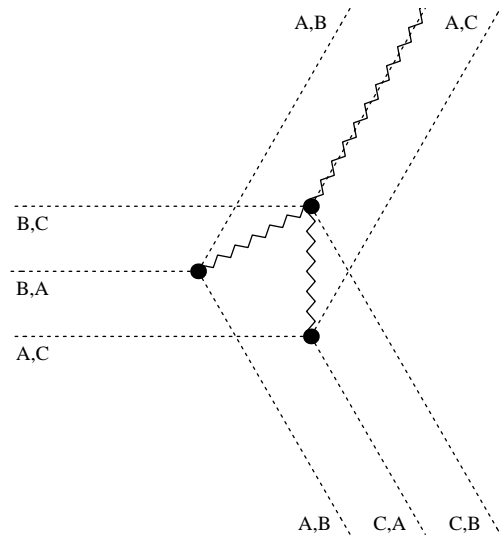}
\caption{\label{fig:stokes1} 
Stokes lines of elementary excitations at a sonic white hole 
with purely imaginary frequencies (dotted lines).
The pairs of letters indicate which branches of the superposition (4.11)
are connected by the lines. 
The first letter of each pair identifies the exponentially dominant branch.}
\end{figure}
The pairs of letters indicate which branches are connected by the lines,
and the first letter identifies the exponentially dominant branch. 
The picture shows that with the choice of branch cuts made we can 
construct a Bogoliubov mode that is acoustic on the lower half plane,
by demanding that $c_{_C}$ vanishes at the C,A Stokes line of the
lowest turning point and that $c_{_B}$ is zero on the lower half plane.
Since $c_{_D}$ is not connected to the other three branches we
can put the coefficient to zero throughout the complex plane.
Such a strategy is not possible for a sonic black hole, see Fig.\ 4, 
because here the order of exponentially dominant or suppressed waves
is exchanged. 
Moreover, we cannot find a solution that is purely acoustic
on the upper half plane, because of the branch cut between 
the central and the highest turning point in Fig.\ 4.
We can of course alter the arrangements of branch cuts, 
such that the cut between the two trans-acoustic branches B and C
points in other directions.
Choosing different branch cuts 
does determine in which half plane the excitation wave 
may be acoustic, but it does not influence whether such an acoustic
behavior is possible at all on either the upper or the lower half plane.
Therefore, at a sonic white hole unstable elementary excitations
with the acoustic asymptotics (\ref{eq:wa}) exists on one of
the complex half planes, 
whereas the unstable modes of
black holes, if any,  are always trans-acoustic.
\begin{figure}
\includegraphics[width=7cm]{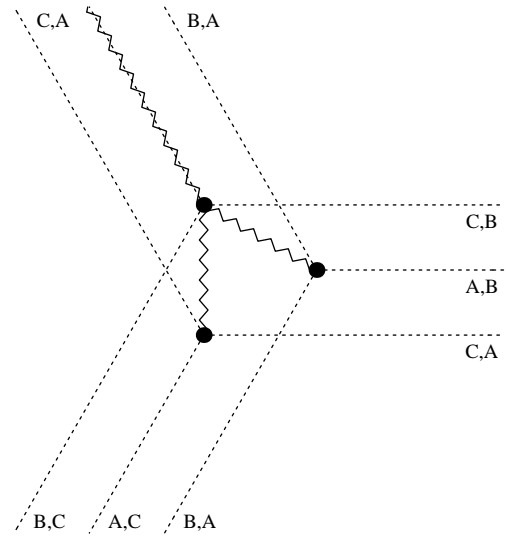}
\caption{\label{fig:stokes2} 
Stokes lines of elementary excitations at a sonic black hole 
with purely imaginary frequencies analogous to Fig.\ 3.}
\end{figure}

\subsection{Instabilities}

Consider the unstable elementary excitations of a sonic white hole,
assuming the acoustic asymptotics (\ref{eq:wa}) 
with complex frequencies $\omega + i\gamma$
to be valid on one of the complex half planes. 
Excitations  (\ref{eq:wa}) with negative $\gamma$ 
are localized near the horizon and are attenuated in time,
whereas acoustic excitations with positive $\gamma$ 
would grow in space and time.
Therefore, the truly unstable modes are trans-acoustic.
Nevertheless, we can use the properties of the attenuated modes
to determine the spectrum of unstable modes, because of the spectral symmetry
of the Bogoliubov-deGennes equations proven in Sec.\ II~C.

Bogoliubov excitations with complex frequencies
have zero norm and should satisfy the orthogonality
relation (\ref{eq:ortho}). 
First we calculate the scalar product
$(w_1,w_2)$ at time $\tau=0$, to find a condition for zero norm.  We
assume that the scalar product (\ref{eq:scalar}) is dominated by the
acoustic region where $|z_0| \ll |z| \ll |c/\alpha|$,
\begin{eqnarray}
 (w_1,w_2) &=& \int w_1^\dagger \sigma_z w_2  \, dz \nonumber \\
&\sim& |A|^2\frac{mc}{\hbar} \left( \int_{- \infty}^0 + \int_0^{+\infty} \right)
\frac{2 \omega}{\alpha z} 
 \,z^{(\omega_1^* - \omega_2)/\alpha} dz \nonumber \\
&=& |A|^2\frac{mc}{\hbar} \left( -e^{\pm 2 \pi \omega/\alpha} + 1\right) \nonumber\\
& & \times \int_0^{\infty} \frac{2\omega}{\alpha z}
 \exp \left( i \frac{ \omega_1^* - \omega_2}{\alpha} \ln z
\right) \, dz \, .
\end{eqnarray}
Here we have connected the two acoustic regions on the upper or lower
half plane, respectively, indicated by the $\pm$ sign, circumventing the
trans-acoustic region close to the horizon at $z=0$.  
The scalar product vanishes if
\begin{equation}
  \omega = 0 \, .
\end{equation}
Therefore, within the validity range of our approximations,
the unstable elementary excitations have purely imaginary
frequencies $\gamma$. 
Let us examine the orthogonality condition
(\ref{eq:ortho}). We represent $\gamma$ as
\begin{equation}
  \label{eq:spectrum}
  \gamma = 2 n \alpha \, ,
\end{equation}
and we deform the integration contour such that it
circumvents the trans-acoustic zone on a large
semi-circle with radium $r$ from below. 
\begin{eqnarray}
  (\overline w _{-n},w_{+n'}) &=& \int \left( v_{-n} u_{+n'} - u_{-n} v_{+n'} \right)
  \, dz \nonumber \\
  &\sim& A_{-n} A_{+n'}\frac{mc}{\hbar} \int \frac{2 n + 2 n'}{ i z} z ^{2
  (n-n')} \, dz \, , \nonumber\\
  &=& A_{-n} A_{+n'}\frac{mc}{\hbar} (2 n + 2 n')\,r^{2 (n-n')} \nonumber\\
  && \times \int_\pi^{2\pi} e^{2i (n-n')\theta} \, d \theta \,.
\end{eqnarray}
The scalar product $( \overline
w_{-n},w_{+n'})$ vanishes for $n \neq n'$ if the $n$ are integers.
The spectrum of unstable elementary excitations consists of discrete
and equally spaced points on the imaginary frequency axis.

An interesting proposal for a sonic hole 
\cite{Garay1,Garay2,Garay3}
involves a toroidal condensate that flows through a constriction where
it exceeds the speed of sound and that then, after the constriction,
settles to subsonic speed.
Garay {\it et al.} \cite{Garay1,Garay2,Garay3}
found that the condensate is unstable only in narrow ``instability
fingers'' in the parameter space used. Our analysis indicates that the
instabilities are generated when the excitations of the toroidal
condensate match the discrete imaginary resonances of the white hole.
This would explain the narrowness of the instability fingers
\cite{Garay1,Garay2,Garay3}. 
Such instabilities are enhanced by the "lasing" effect of
the black-hole white-hole pair \cite{CJ}.
The white hole generates a hydrodynamic instability that is
resonantly enhanced by the pair of horizons 
where elementary excitations can bounce back and fourth.
Sonic black holes can be stabilized by employing the equivalent
of a Laval nozzle \cite{Laval} that converts a subsonic flow
to a supersonic one without causing turbulence 
(as in a rocket engine).
Our theory indicates that white holes are intrinsically unstable
\cite{LKOWhite}, generating breakdown shocks \cite{Courant}.

\section{Summary}

Unstable Bose-Einstein condensates develop elementary excitations with
complex frequencies. Such excitations have zero norm and are subject
to orthogonality relations between pairs of excitations with opposite
frequencies. We elaborated the general theory of unstable elementary
excitations and of two important approximate methods to analyze their
behaviour, the acoustic and the WKB approximations. 
Applying these techniques, we showed that sonic white holes in
Bose-Einstein condensates give rise to a discrete spectrum of
instabilities, which may explain the remarkable stability of sonic
holes in toroidal traps \cite{Garay1,Garay2,Garay3}.

\section*{ACKNOWLEDGEMENTS}

We thank J. R. Anglin, M. V. Berry, I. Bialynicki-Birula,
I. A. Brown, L. J. Garay, 
T. A. Jacobson, R. Parentani, S. Stenholm, M. Visser, and G. E. Volovik
for discussions.
Our work was supported by the ESF Programme 
Cosmology in the Laboratory,
the Leverhulme Trust,
the National Science Foundation of Hungary (contract No.\ F032346),
the Marie Curie Programme of the European Commission,
the Royal Society of Edinburgh,
and by the Engineering and Physical Sciences Research Council.

\end{document}